\documentclass[prl,twocolumn,superscriptaddress,showpacs]{revtex4-1}
\usepackage[dvips]{graphicx,color}
\usepackage{amsmath, amssymb, amsxtra}
\usepackage{bm,array}
\def\U#1{{\rm #1}} 
\def\u#1{_{\rm #1}}
\newcommand{\ket}[1]{| #1 \rangle}
\newcommand{\bra}[1]{\langle #1 |}
\newcommand{\ketbra}[2]{| #1 \rangle \langle #2 |}
\newcommand{\vect}[1]{\boldsymbol{#1}}

\begin{document}
\title{
Polarization insensitive frequency conversion for an atom-photon entanglement distribution via a telecom network
}

\author{Rikizo~Ikuta}
\affiliation{Graduate School of Engineering Science, Osaka University,
  Toyonaka, Osaka 560-8531, Japan}
\author{Toshiki~Kobayashi}
\affiliation{Graduate School of Engineering Science, Osaka University,
  Toyonaka, Osaka 560-8531, Japan}
\author{Tetsuo~Kawakami}
\affiliation{Graduate School of Engineering Science, Osaka University,
  Toyonaka, Osaka 560-8531, Japan}
\author{Shigehito~Miki}
\affiliation{Advanced ICT Research Institute, 
National Institute of Information and Communications Technology (NICT),
Kobe 651-2492, Japan}
\author{Masahiro~Yabuno}
\affiliation{Advanced ICT Research Institute, 
National Institute of Information and Communications Technology (NICT),
Kobe 651-2492, Japan}
\author{Taro~Yamashita}
\affiliation{Advanced ICT Research Institute, 
National Institute of Information and Communications Technology (NICT),
Kobe 651-2492, Japan}
\affiliation{
Japan Science and Technology Agency, PRESTO, Kawaguchi, Saitama
332-0012, Japan}
\author{Hirotaka~Terai}
\affiliation{Advanced ICT Research Institute, 
National Institute of Information and Communications Technology (NICT),
Kobe 651-2492, Japan}
\author{Masato~Koashi}
\affiliation{Photon Science Center, Graduate School of Engineering,
  The University of Tokyo, Bunkyo-ku, Tokyo 113-8656, Japan}
\author{Tetsuya~Mukai}
\affiliation{NTT Basic Research Laboratories, 
  NTT Corporation, Atsugi, Kanagawa 243-0198, Japan}
\author{Takashi~Yamamoto}
\affiliation{Graduate School of Engineering Science, Osaka University,
  Toyonaka, Osaka 560-8531, Japan}
\author{Nobuyuki~Imoto}
\affiliation{Graduate School of Engineering Science, Osaka University,
  Toyonaka, Osaka 560-8531, Japan}

\begin{abstract}
  Quantum network with a current telecom photonic infrastructure is deficient 
  in quantum storages that keep arbitrary quantum state in sufficient time duration 
  for a long-distance quantum communication with quantum repeater algorithms. 
  Atomic quantum storages have achieved subsecond storage time
  corresponding to 1000~km transmission time for a telecom photon through a quantum repeater algorithm. 
  However, the telecom photon is not directly accessible to typical atomic storages.
  Solid state quantum frequency conversions fill this wavelength gap 
  and add more abilities, for example, a frequency multiplex. 
  Here we report on the experimental demonstration of a polarization-insensitive 
  solid-state quantum frequency conversion to a telecom photon 
  from a short-wavelength photon entangled with an atomic ensemble. 
  Atom-photon entanglement has been generated with a Rb atomic ensemble 
  and the photon has been translated to telecom range 
  while retaining the entanglement
  by our nonlinear-crystal-based frequency converter in a Sagnac interferometer. 
\end{abstract}

\maketitle

Quantum frequency conversion~\cite{Kumar1990}~(QFC)
based on nonlinear optical processes enables us
to change the color of photons without destroying the quantum properties. 
This allows us to transfer quantum properties of a physical system 
to another one which have different accessible frequencies 
through a single photon~\cite{Tanzilli2005,Dudin2010,McGuinness2010,Rakher2010,Ikuta2011,Ramelow2012,Zaske2012,Ates2012,De2012,Ikuta2013,Rutz2017,Liu2017}. 
Besides that, 
we can use QFC for other purposes such as 
erasing distinguishability of photons~\cite{Takesue2008},
manipulating spectral and temporal modes of photons~\cite{Kielpinski2011,Brecht2011,Lavoie2013,Fisher2016,Matsuda2016,Manurkar2016,Kroh2017,Allgaier2017,Allgaier2017-2},
and performing frequency-domain quantum information processing~\cite{Kobayashi2016,Clemmen2016,Kobayashi2017} 
by tailoring of the pump light. 
Most of those abilities have been demonstrated with solid-state QFC devices 
because of its applicability to a wide frequency range, 
which is similar to the mirrors and beamsplitters~(BSs) 
for the spatial manipulation of the photons. 

The extension of the solid-state QFC 
for the quantum storages have also been actively studied~\cite{Fernandez2013,Farrera2016,Ikuta2016}.
For a long-distance quantum communication, 
a long lifetime quantum storage that entangled with a telecom photon is necessary.
The cold Rb atomic ensemble is one of the promising quantum storage 
that has a long lifetime and a high efficiency atom-photon entanglement generation~\cite{Yuan2008,Zhao2009,Dudin2010,Ritter2012,Bao2012,Hofmann2012,Yang2016}. 
Recently, solid-state QFC of a single photon from the cold Rb atomic ensemble 
has been demonstrated~\cite{Farrera2016, Ikuta2016}, 
but the quantum state preservation has never been shown yet.
In this letter, we first describe our polarization insensitive solid-state QFC 
by a waveguided periodically-poled lithium niobate~(PPLN) installed in a Sagnac interferometer. 
Then we show creation of entanglement between the Rb atomic state and a telecom photon. 

\section*{Results}
{\bf Polarization-insensitive QFC.} 
We first review the conventional QFC of a single-mode light
with a specific polarization 
based on the second-order nonlinear optical effect ~\cite{Kumar1990,Ikuta2011}. 
When a pump light at angular frequency $\omega\u{p}$ is sufficiently strong, 
the Hamiltonian of the process is 
described by $H=i\hbar \xi^* a^\dagger\u{l}a\u{u}+\U{h.c.}$, 
where h.c. represents the Hermitian conjugate of the first term, 
and $a\u{u}$ and $a\u{l}$ are annihilation operators 
of upper and lower frequency modes at angular frequencies 
$\omega\u{u}$ and $\omega\u{l}(=\omega\u{u}-\omega\u{p})$, respectively. 
Coupling constant $\xi=|\xi|e^{i\phi}$ is proportional to 
the complex amplitude of the pump light with its phase $\phi$. 

\begin{figure*}[t]
 \begin{center}
 \scalebox{1.0}{\includegraphics{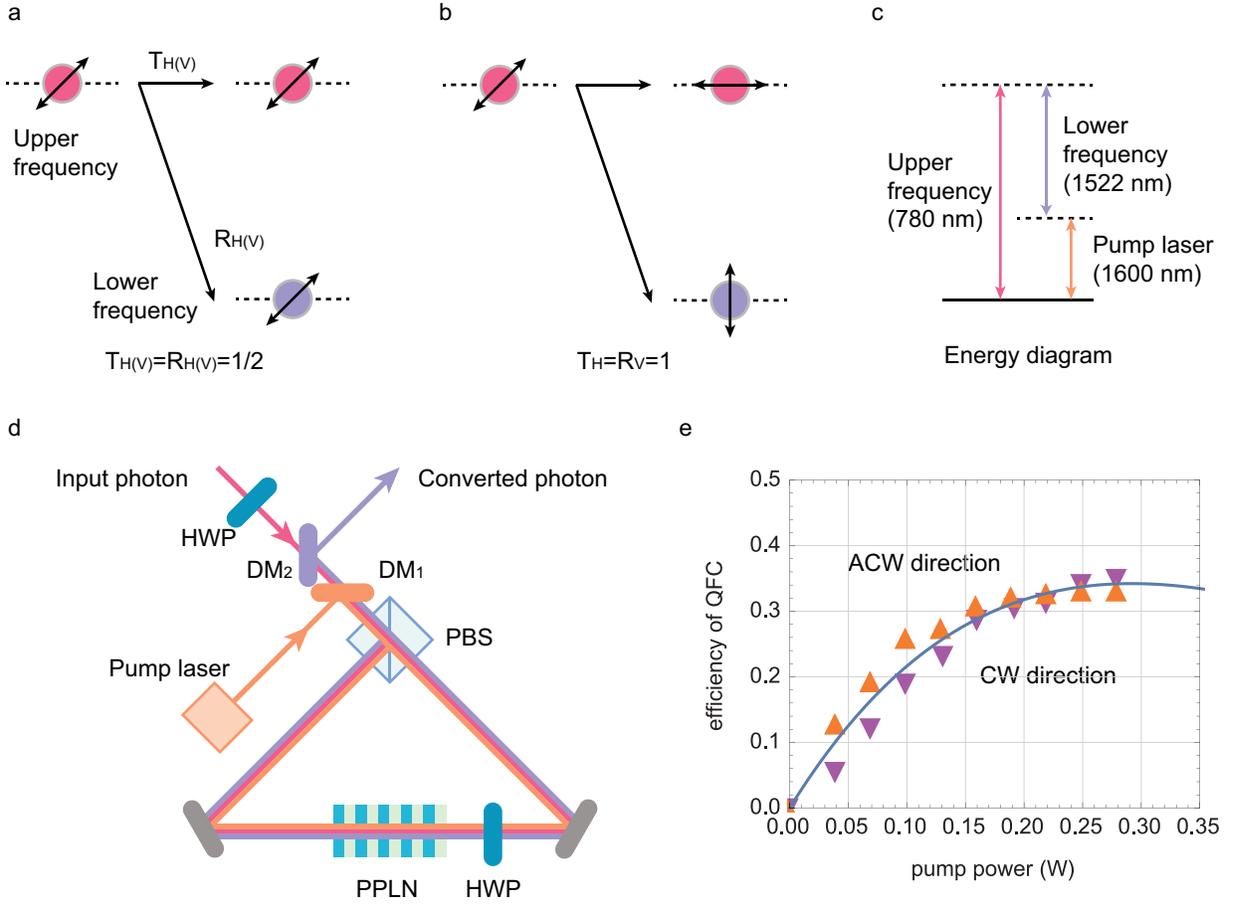}}
  \caption{
  {\bf QFC as a frequency-domain linear optics with the two polarization modes.}
  {\bf a},
  Concept of a non-polarizing frequency-domain half BS 
  for $T\u{H(V)} = R\u{H(V)} = 1/2$. 
  {\bf b},
  A frequency-domain PBS 
  for $T\u{H}=R\u{V}=1$ and $T\u{V}=R\u{H}=0$. 
  {\bf c},
  Energy diagram related to the QFC used in our experiment. 
  The difference frequency generation of a photon 
  from 780~nm~(upper frequency)
  to 1522~nm~(lower frequency) is
  performed by using a strong pump light at 1600~nm. 
  {\bf d},
  The experimental setup of dual-polarization-mode QFC.
  Type-0 quasi-phase-matched PPLN crystal as a nonlinear optical medium 
  which acts on only the V-polarized photons 
  is installed in the Sagnac type interferometer.
  The detailed explanation of the QFC is in the main text. 
  {\bf e},
  Conversion efficiency 
  from the front to the end of the QFC. 
  It is measured by using classical laser light and a power meter.
  A solid curve is fitted by using a function $\eta\u{max}\sin^2(\sqrt{g P})$,
  where $P$ is the pump power.
  Parameters $\eta\u{max}$ and $g$ are estimated as 0.34 and 8.4/W, respectively.  
  \label{fig:qfc}}
 \end{center}
\end{figure*}
As shown in Fig.~\ref{fig:qfc}d, when two QFCs working 
for two different polarization modes are superposed, 
the interaction Hamiltonian can be described by 
$H = i\hbar (\xi^*\u{H} a^\dagger\u{l,H} a\u{u,H}+\xi^*\u{V} a^\dagger\u{l,V}a\u{u,V})+\U{h.c.}$,
where $a\u{u(l),H}$ and $a\u{u(l),V}$ are annihilation operators 
of horizontally(H-) polarized and vertically(V-) polarized upper(lower) frequency modes,
respectively. 
$\xi\u{H(V)}=|\xi\u{H(V)}|e^{i\phi\u{H(V)}}$ is proportional to 
the amplitudes of the H(V)-polarized pump light with phase $\phi\u{H(V)}$. 
By using the Heisenberg representation, 
annihilation operators $a\u{u,H(V),out}$ and $a\u{l,H(V),out}$ of 
the upper and lower frequency modes coming from the nonlinear optical medium are 
represented by 
\begin{eqnarray}
  a\u{u,H(V),out}= t\u{H(V)} a\u{u,H(V)} - r\u{H(V)}a\u{l,H(V)}
  \label{eq:s}
\end{eqnarray}
and
\begin{eqnarray}
  a\u{l,H(V),out}=  r^*\u{H(V)}a\u{u,H(V)} + t\u{H(V)} a\u{l,H(V)}, 
  \label{eq:i}
\end{eqnarray}
where 
$t\u{H(V)}=\cos(|\xi\u{H(V)}|\tau)$ and 
$r\u{H(V)}=e^{i\phi\u{H(V)}} \sin(|\xi\u{H(V)}|\tau)$. 
$\tau$ is the travelling time of the light pulses through the medium. 
The transmittance $T\u{H(V)}\equiv |t\u{H(V)}|^2$ and 
the reflectance $R\u{H(V)}\equiv |r\u{H(V)}|^2$ 
can be changed by adjusting the amplitudes of the pump light. 
When $T\u{H}=T\u{V}$ and $R\u{H} = R\u{V}$ are satisfied,
and a single photon converted event is postselected, 
the QFC process while preserving the input polarization state 
up to the constant phase shift of $\phi\u{H}-\phi\u{V}$ 
is achieved. The phase shift can be compensated in principle, 
and in that sense, we call this process 
the polarization insensitive QFC. 
Based on equations~(\ref{eq:s}) and (\ref{eq:i}), 
this dual-polarization-mode QFC has additional interesting features. 
(a) For $T\u{H(V)} = R\u{H(V)} = 1/2$,
it becomes a non-polarizing frequency-domain half BS~\cite{Kobayashi2016,Clemmen2016,Kobayashi2017,Ikuta2014}~(See Fig.~\ref{fig:qfc}a). 
(b) For $T\u{H}=R\u{V}=1$ and $T\u{V}=R\u{H}=0$, 
it becomes a frequency-domain polarizing BS~(PBS)~(See Fig.~\ref{fig:qfc}b). 
(c) For $T\u{H} \neq T\u{V}$ and $0 < T\u{H(V)} < 1$, 
it becomes a frequency-domain partially-polarizing BSs~(PPBS). 
PPBSs with proper settings of the transmittance and the reflectance 
can be used to perform frequency-domain quantum information protocols 
such as entanglement distillation~\cite{Bennett1996,Kwiat2012}, 
probabilistic nonlinear optical gate~\cite{Knill2000,Okamoto2011},
quantum state estimation~\cite{Ling2006} 
and manipulation of multipartite entangled states~\cite{Tashima2009,Ikuta2011-2}. 

We explain the experimental detail of the polarization insensitive QFC
in Fig.~\ref{fig:qfc}d. 
The nonlinear optical medium for QFC is 
a type-0 quasi-phase-matched PPLN waveguide which 
converts a V-polarized input photon to a V-polarized photon 
with the use of the V-polarized pump light. 
The PPLN is installed in a Sagnac interferometer.
In this demonstration, 
we prepare a polarizing upper frequency photon at 780~nm 
entangled with a Rb atomic ensemble 
as an input signal to the converter which we explain in detail later. 
As shown in Fig.~\ref{fig:qfc}c,
by using a strong pump light at 1600~nm with a linewidth of 150~kHz, 
the upper frequency photon at 780~nm is 
converted to the lower frequency photon at 1522~nm 
by difference frequency generation~(DFG). 
In Fig.~\ref{fig:qfc}d, the polarization of the input photon is 
flipped from H~(V) polarization to V~(H)
by a half wave plate~(HWP). 
The photon is combined with the diagonally polarized strong pump light at 1600~nm 
at a dichroic mirror~(DM1). 
At a PBS, the H- and V-polarized components of them 
are split into clockwise~(CW) and anti-clockwise~(ACW) directions, respectively. 
For the CW direction, 
after flipped from H to V polarization at a HWP, 
the V-polarized input photon and the pump light are coupled to the PPLN waveguide.
After the conversion, 
the V-polarized photons and pump light are reflected by the PBS. 
Then, only the converted photon is extracted from the reflection port of DM2, 
being separated from the pump light 
and the residual input photon by DM1 and DM2, respectively. 
On the other hand, for the ACW direction, 
the V-polarized input photon and the pump light are coupled to the PPLN waveguide. 
After the conversion, the polarization is flipped from V to H by the HWP, 
and the photons and the pump light pass through the PBS. 
Finally, only the converted photon is extracted by DM1 and DM2.
The conversion efficiencies of the QFC for CW and ACW directions
are shown in Fig.~\ref{fig:qfc}e. 

{\bf Experimental setup of entanglement between atoms and a telecom photon.} 
\begin{figure*}[t]
 \begin{center}
 \scalebox{1}{\includegraphics{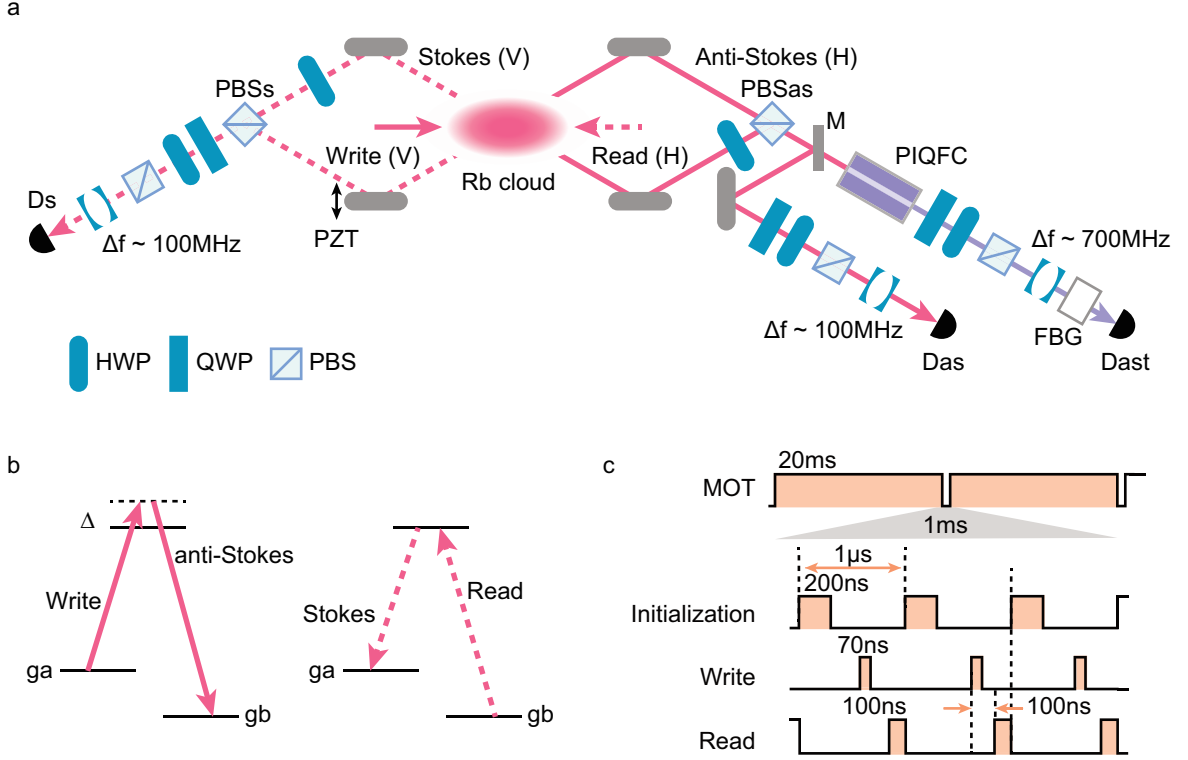}}
  \caption{
  {\bf Experimental setup.}
  {\bf a},
  Our experimental setup for entanglement between Rb atoms
  and visible/telecom photons without/with QFC. 
  When mirror M is flipped up, AS photon is detected
  by $\U{D}\u{as}$ without QFC. 
  When mirror M is flipped down, AS photon is input 
  to polarization-insensitive QFC~(PIQFC) in Fig.~\ref{fig:qfc}d 
  and then the converted photon is detected by $\U{D}\u{ast}$. 
  In order to stabilize the interferometer in optical paths 
  of Stokes and anti-Stokes photons, 
  we use a conventional diode laser at a center wavelength of 850~nm~(not shown). 
  The light enters into the interferometer from the vaccum port of $\U{PBS}\u{s}$, 
  and then after passing along the same four optical paths 
  for Stokes and anti-Stokes photons, 
  it comes from the vaccum port of $\U{PBS}\u{as}$. 
  The light is detected by a photo detector, and
  its detection signal is used for feedback control
  of the interferometer by a mirror on a Piezo stage~(PZT). 
  {\bf b},
  $\Lambda$-type energy levels of $\U{D}_2$ line in $^{87}\U{Rb}$ used
  in our experiment. The ground levels of $g\u{a}$ and $g\u{b}$ correspond to
  the levels of $5^{2}S_{1/2}, F=2$ and $5^{2}S_{1/2}, F=1$, respectively. 
  The excited level is $5^{2}P_{3/2}, F'=2$.
  The magnetic sublevel is degenerated in our experiment. 
  {\bf c},
  Time sequence of the experiment. 
  The quantum experiment is performed within 1~ms 
  during the MOT is turned off.
  The period of the two initialization pulses is 1~$\mu$s and 
  the injection of them is repeated 990 times within 1~ms. 
  \label{fig:setup}}
 \end{center}
\end{figure*}
In order to prepare a 780-nm signal photon entangled with the Rb atoms, 
we construct an experimental setup as shown in Fig.~\ref{fig:setup}a. 
We use $\Lambda$-type energy levels of $\U{D}_2$ line at 780~nm 
in $^{87}\U{Rb}$ atoms 
($5^{2}S_{1/2}\leftrightarrow 5^{2}P_{3/2}$) as shown in Fig.~\ref{fig:setup}b. 
We prepare the Rb atomic ensemble by a magneto-optical trap~(MOT) 
in 20~ms~(see Fig.~\ref{fig:setup}c). 
After the trapping lasers and the magnetic field for the MOT are turned off, 
we perform the QFC experiment 990 times within 1~ms. 
A horizontally~(H-) polarized 200-ns initialization pulse 
at the resonant frequency between a ground level $g\u{b}$~($F=2$) 
and an excited level~($F'=2$) initializes the atoms 
into another ground level $g\u{a}$~($F=1$). 
Then a vertically~(V-) polarized 70-ns write pulse 
blue-detuned by $\Delta \sim$10~MHz from the resonant frequency
between $g\u{a}$ and the excited level is injected to the atoms,
causing the Raman transition 
from $g\u{a}$ to $g\u{b}$ with emission of anti-Stokes~(AS) photons. 

The momentum conservation guarantees that 
the wave vector $\vect{k}\u{atom}$ of the collective spin excitation of the atoms
satisfies $\vect{k}\u{atom}=\vect{k}\u{W}-\vect{k}\u{AS}$, 
where $\vect{k}\u{W/AS}$ is the wave vector of the write/AS light. 
When we postselect a particular wave vector~(path) of 
a single AS photon in a single mode
whose quantum state is denoted by $\ket{path_x}\u{AS}$, 
the wave vector of the atoms corresponding to the photonic state 
is decided by the momentum conservation. 
We denote the atomic state by $\ket{k_{x}}\u{atom}$. 
As a result, when we postselect two different wave vectors of the AS photon 
corresponding to the states $\ket{path_+}\u{AS}$ and $\ket{path_-}\u{AS}$, 
we obtain a quantum state of the atoms and the AS photon as 
\begin{eqnarray}
\alpha\ket{path_+}\u{AS}\ket{k_+}\u{atom}+\beta\ket{path_-}\u{AS}\ket{k_-}\u{atom}, 
\label{eq:entanglement}
\end{eqnarray}
with $|\alpha|^2+|\beta|^2=1$.
The subscripts $+$ and $-$
imply the upper and lower optical paths of the AS photons in Fig.~\ref{fig:setup}a. 
By adjusting the excitation probabilities such that $\alpha=\beta$ is satisfied, 
we obtain a maximally entangled state. 
In our experiment,
we select the H-polarized AS photons emitted in two directions 
at small angles~($\sim \pm 3^\circ$) relative to the direction of the write pulse, 
whose emission probabilities are the same ideally. 
By using a HWP and a PBS~($\U{PBS}\u{as}$), 
path information of AS photons is transformed to polarization information. 
After the operation, 
$\ket{path_{+/-}}\u{AS}$ in Eq.~(\ref{eq:entanglement}) 
is changed to the H/V-polarized states denoted by $\ket{H/V}\u{AS}$,
and we obtain 
\begin{eqnarray}
(\ket{H}\u{AS}\ket{k_+}\u{atom}+\ket{V}\u{AS}\ket{k_-}\u{atom})/\sqrt{2}. 
\label{eq:entanglement2}
\end{eqnarray}

In order to evaluate the quantum correlation between the atoms and the photons, 
we inject an H-polarized 100-ns read light at the resonant frequency 
between $g\u{b}$ and the excited level into the atoms. 
The read light provides the transition of the Rb atoms to 
$g\u{a}$ and generation of the Stokes~(S) photons.
In our experiment, we collect only the V-polarized component of the S photons. 
Because of the momentum conservation, 
wave vector $\vect{k}\u{S}$ of the S photons 
satisfies $\vect{k}\u{S}=\vect{k}\u{atom}+\vect{k}\u{R}$. 
The direction of the emitted S photons is 
decided by the wave vector of the atomic excitation. 
Because such a read operation does not access the AS photon, 
the operation never increase or newly create the entanglement
between the atoms and the AS photon. 
Thus observation of an entangled state of 
the which-path state of the S photon and the polarizing AS photon
after the read operation 
is the evidence of the entanglement
between the atoms and the AS photon before the read operation. 

In the experiment, 
we inject the read pulse from the direction opposite to the write pulse. 
The wave vector $\vect{k}\u{R}$ of the read pulse satisfies 
$\vect{k}\u{R} \sim - \vect{k}\u{W}$, 
leading to the relation 
$\vect{k}\u{S}\sim -\vect{k}\u{AS}$ from the momentum conservation. 
This means the S photons are emitted in a direction at $\sim \mp 3^\circ$ 
relative to the direction of the read pulse 
when the AS photons are emitted 
in a direction at $\sim \pm 3^\circ$ relative to that of the write pulse. 
By using a HWP and a PBS~($\U{PBS}\u{s}$) shown in Fig.~\ref{fig:setup}~a, 
the path information of the V-polarized S photons is 
transformed into the polarization. 
Finally, we can observe the entanglement between the atoms and the AS photons 
through the polarization entangled photon pair of the AS and the S photons. 

After a polarization analyser 
composed of a QWP, a HWP and a PBS for the quantum state tomography~\cite{2001James}, 
S photon passes through 
a monolithic cavity-coated lens as a frequency filter 
with a bandwidth of $\sim$100~MHz~\cite{Palittapongarnpim2012} 
and is coupled to a single-mode optical fiber. 
Then S photon is detected 
by a silicon avalanche photon detector~(APD) denoted by $\U{D}\u{s}$
with a quantum efficiency of $\sim 60$~\%. 

When we do not perform QFC, 
AS photon is detected by another APD~($\U{D}\u{as}$) 
after passing through a polarization analyzer, 
a cavity-coated lens with a bandwidth of $\sim$100~MHz 
and a single-mode optical fiber. 
When we perform QFC, 
mirror M in Fig.~\ref{fig:setup}a is flipped down 
in order to send the AS photon to the polarization-insensitive QFC 
depicted in Fig.~\ref{fig:qfc}d. 
We set the conversion efficiency from the front of the PPLN waveguide 
to the end of the QFC to $\sim 30$~\%
by using the effective pump power of $\sim 0.2$~W~(see Fig.~\ref{fig:qfc}e). 
The telecom photon from the QFC passes through a polarization analyser 
followed by an etalon with a bandwidth of $\sim$700~MHz, 
and a pair of fiber Bragg gratings with a total bandwidth of $\sim$0.8~GHz. 
Finally, the telecom photon is detected by a superconducting single photon detector~(SSPD)
denoted by $\U{D}\u{ast}$ with a quantum efficiency of $\sim 60$~\%~\cite{Miki2017}. 

We repeat the above measurement about 47000 times per second. 
We use a trigger signal for starting each sequence as 
a start signal of a time-to-digital converter~(TDC). 
The photon countings measured by $\U{D}\u{s}$, $\U{D}\u{as}$ and $\U{D}\u{ast}$
are used as stop signals of the TDC. 
We collect the coincidence events 
between the signals of modes S and AS in their time windows of 64~ns. 

{\bf Experimental results.} 
\begin{figure}[t]
 \begin{center}
  \scalebox{0.55}{\includegraphics{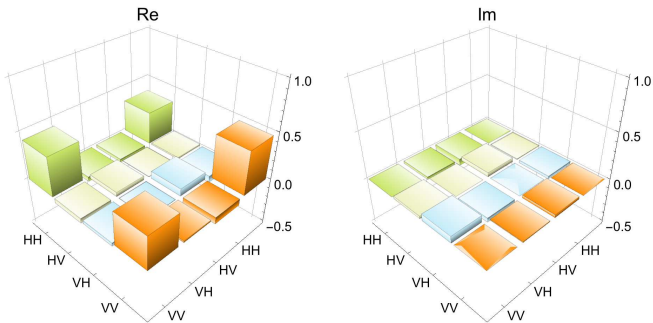}}
  \caption{
    The reconstructed density matrix 
    $U_{\theta_0}^\dagger \rho\u{S,AS} U_{\theta_0}$ without QFC. 
  \label{fig:without}}
 \end{center}
\end{figure}
Without QFC, 
we performed the quantum state tomography between the S photon and the AS photon,
and reconstructed density operator $\rho\u{S,AS}$ 
by the use of the iterative maximum likelihood method~\cite{2007Rehacek}. 
We estimated entanglement of formation~\cite{Wootters1998}~(EoF) $E$ and 
the purity defined by $P = \U{tr}(\rho\u{S,AS}^2)$ as
$E = 0.37 \pm 0.11$ and $P=0.61 \pm 0.06$, respectively. 
We also estimated a maximized fidelity to a maximally entangled state 
$U_\theta\ket{\phi^+}$ 
defined by $F=\U{max}_\theta \bra{\phi^+}U_\theta^\dagger \rho\u{S,AS}U_\theta\ket{\phi^+}$,
whose value was $F = 0.78\pm 0.05$ for $\theta=\theta_0=-65^\circ$.
Here
$\ket{\phi^+}=(\ket{H}\u{AS}\ket{H}\u{S}+\ket{V}\u{AS}\ket{V}\u{S})/\sqrt{2}$ 
and 
$U_\theta = \exp(-i\theta Z/2)\otimes I$ 
with $Z=\ketbra{H}{H}-\ketbra{V}{V}$ and $I=\ketbra{H}{H}+\ketbra{V}{V}$. 
The matrix representation of density operator 
$U_{\theta_0}^\dagger \rho\u{S,AS} U_{\theta_0}$ is shown in Fig.~\ref{fig:without}. 
These results show the entanglement between AS photon and the Rb atoms. 
The observed count rate of the two-photon state was about $0.08$~Hz
through the overall experiment time of 16~hours
including the load time of the atoms by MOT. 

\begin{figure}[t]
 \begin{center}
  \scalebox{0.55}{\includegraphics{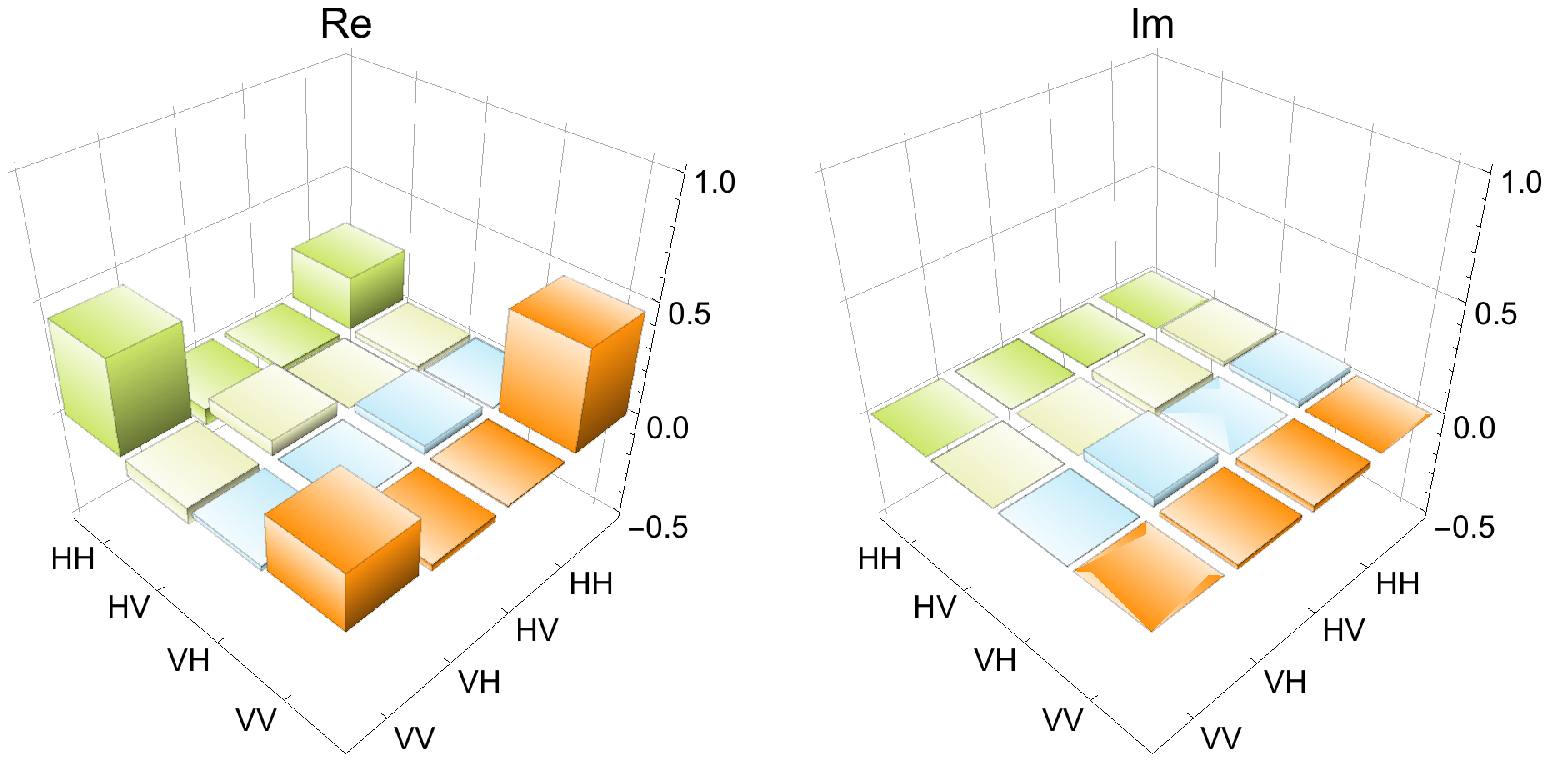}}
  \caption{
    The reconstructed density matrix $U_{\theta_1}^\dagger \rho\u{S,ASt} U_{\theta_1}$
    with QFC.
  \label{fig:with}}
 \end{center}
\end{figure}
With QFC,
we performed the quantum state tomography 
between S photon and the wavelength-converted AS photon. 
The estimated EoF and purity of 
reconstructed density operator $\rho\u{S,ASt}$ 
were $E = 0.25 \pm 0.13$ and $P = 0.55 \pm 0.07$, respectively. 
The maximized fidelity to $U_\theta\ket{\phi^+}$ about $\theta$ 
was $F = 0.69\pm 0.07$ for $\theta=\theta_1=93^\circ$. 
The matrix representation of density operator 
$U_{\theta_1}^\dagger \rho\u{S,ASt}U_{\theta_1}$ is shown in Fig.~\ref{fig:with}. 
The EoF $E$ is clearly greater than $0$, 
which shows that the state of the Rb atoms and the telecom photon has entanglement. 
From the result, we succeeded the creation of the entanglement 
between the Rb atoms and the telecom photon 
by using the polarization-insensitive QFC. 
The observed count rate of the two-photon state was about $0.0065$~Hz 
through the experiment time of 83 hours. 

\section*{Discussion}
In conclusion, 
we have shown the entanglement between
the wave vector of the collective spin excitation of the Rb atoms 
and the polarizing telecom photon 
by using the polarization insensitive QFC 
composed of the PPLN waveguide installed in a Sagnac interferometer. 
Recent researches showed the efficient
and sub-second lifetime quantum memory~\cite{Yang2016} 
and a multiplexed quantum memory~\cite{Pu2017,Pu2017-2} by using Rb atomic ensembles. 
Combining such state-of-the-art technologies with our experimental result 
will be useful for fiber-based quantum communication over long distance.
Furthermore, the demonstrated polarization insensitive QFC is applicable 
to various conversion systems for matter-based quantum storages,
such as trapped ions~\cite{Lenhard2017,Krutyanskiy2017},
diamond color centers~\cite{Ikuta2014},
and quantum dots~\cite{De2012},
which are useful for measurement-based quantum computation~\cite{Briegel2009,Ladd2010,Buluta2011}.
The potential of our QFC is not limited by the use of 
the polarization insensitive QFC. 
The proposed dual-polarization-mode QFC 
has additional features as 
a non-polarizing frequency-domain half BS, 
a frequency-domain PBS and a frequency-domain PPBS. 
The devices will provide various kinds of tasks developed 
in the linear optical quantum information processing. 

\section*{Acknowledgements}
We thank Yoshiaki Tsujimoto and Motoki Asano for helpful discussions about QFC. 
This work was supported by 
CREST, JST JPMJCR1671;
MEXT/JSPS KAKENHI Grant Number JP26286068, JP15H03704, JP16H02214 and JP16K17772.

\end{document}